\documentclass[letter]{aa}

\usepackage[utf8]{inputenc}

\usepackage{newtxtext,newtxmath}
\usepackage{graphicx}
\usepackage{xcolor}
\usepackage{subfigure}
\usepackage{bm}
\usepackage{ulem}
\usepackage{mathtools}
\usepackage{amsmath}
\usepackage{lastpage}

\def\me{$\,{\rm M}_{\oplus}\,$}
\def\re{$\,{\rm R}_{\oplus}\,$}

\definecolor{mygray}{gray}{0.6}
\definecolor{orange}{rgb}{1.0, 0.4, 0.0}

\usepackage{ulem}

\begin{document} 

\title{Rocky sub-Neptunes formed by pebble accretion:\\
Rain of rocks from polluted envelopes}
\titlerunning{Rainout in rocky sub-Neptunes formed by PA}

\author{Allona Vazan\inst{1} \and Chris W. Ormel\inst{2}}
\institute{Astrophysics Research Center (ARCO), Department of Natural Sciences, The Open University of Israel, Raanana 4353701, Israel \and 
Department of Astronomy, Tsinghua University, 30 Shuangqing Rd, 100084 Beijing, China}
\date{November 2022}

\abstract
{Sub-Neptune planets formed in the protoplanetary disk accreted hydrogen-helium (H,He) envelopes. Planet formation models of sub-Neptunes formed by pebble accretion result in small rocky cores surrounded by polluted H,He envelopes where most of the rock (silicate) is in vapor form at the end of the formation phase. This vapor is expected to condense and rain-out as the planet cools. 
In this Letter we examine the timescale for the rainout and its effect on the thermal evolution. 
We calculate the thermal and structural evolution of a 10\me planet formed by pebble accretion, taking into account material redistribution {from} silicate rainout (condensation and settling) and from convective mixing. 
We find that the {duration of the} rainout in sub-Neptunes {is on {${\sim}$Gyr} timescale and} varies with envelope mass: planets with {envelopes below} $\sim$0.75\me rainout into a core-envelope structure in less than {$1\,\mathrm{Gyr}$}, while planets {in excess of} 0.75\me {of} H,He preserve some of their envelope pollution for billions of years. 
The energy released by the rainout {inflates the radius with respect to planets that start out from {a} plain core-envelope structure. This inflation} {would result in} estimates of the H,He contents of observed exoplanets based on the standard core-envelope structure {to be too high}. 
{We identify a number of planets in the exoplanet census where rainout may operate, which would result in their H,He contents to be overestimated by up to a factor two.} Future accurate age measurements by the PLATO mission may allow the identification of planets formed with polluted envelopes.}

\maketitle

             
\section{Introduction}

The radius valley \citep{fulton17} 
splits the population of close-in exoplanets into two groups: bare rocky objects, and larger radii planets. This {dichotomy} may distinguish dry super-Earths from sub-Neptunes, which contains lighter materials - volatiles and/or hydrogen and helium (hereafters H,He).  
While the interior structure of super-Earth planets can be {modeled} to some degree from the Earth's structure, the structure of sub-Neptunes is yet unclear. 
Based on the detected radii {and masses}, close-in sub-Neptunes contain only a few percent of H,He / volatile. Planet formation models predict more gas for planets in this mass range \citep{leechiang15,mordasini20,ormel21}, where mass loss processes after disk dissipation removed some of the gas \citep{owenwu13,gupta20,rogersowen21}. If so, farther-out sub-Neptune planets may have thicker envelopes than the close-in planets we observe today. 
    
Planet formation models of rocky sub-Neptune planets show that solid {sublimation} in the growing gas envelope leads to massive pollution of the envelope with silicate vapor once a planet exceeds about 2\me \citep{boden18,brouwers18,ormel21}. 
The formation of sub-Neptunes by rocky pebble accretion results in a typical structure of a small rocky core surrounded by a silicate vapor composition gradient and a vapor-rich convective envelope on top of it \citep{ormel21}. 
Progressive cooling of these planets lead to over-saturation of the silicate vapors in the envelope and consequently condensation and settling (rainout) of rock droplets to deeper undersaturated layers. Rainout sweeps the silicate from the outer layers downwards and enhances late core growth \citep{brouwers20}. 
{The} Rainout phase {terminates} in a core-envelope structure, where further cooling won't change the interior structure. The timescale for rainout depends on the thermal evolution of the planet. \cite{brouwers20} estimated the timescales for rainout in super-Earths to be billions of years, calculated for uniformly polluted envelopes, based on ideal gas and adiabatic cooling scheme. 
Other recent works explored implications of hydrogen-rock distribution in the interior of sub-Neptune planets, in which the structure doesn't evolve into a core-envelope structure \citep{markham22,misener22}, or that rainout process wasn't {explicitly} modeled \citep{boden18}.

In this Letter we {connect planet formation with evolution - we} {numerically} calculate {the post-formation} evolution of a 10\me sub-Neptune formed by rocky pebble accretion. We consider {a} {simultaneous} 
thermal and structural evolution of the interior, which at formation has a silicate composition gradient and a polluted envelope. In particular, we follow the rainout process and its consequences {for} {the radius evolution and mass-radius relation.} 
The 10\me case was selected as an intermediate case to {emphasize} the difference between interiors of smaller super-Earths and larger Neptunes. 
A companion paper describes the evolution of 5-20\me planets formed by rocky pebble accretion under various conditions (Vazan et al. - in prep.). 
    
\section{Method}\label{sec:method}

In this section we describe the new methods used in this work which build upon the interior evolution methods and results of previous {works} \citep{vazan15,vazan18c,vazan22}.

{\it The initial model} is the outcome of planet formation model of rocky pebble accretion \citep{ormel21}. The formation model was calculated for {a} solid accretion rate {of} $10^{-5}${\me\,yr$^{-1}$} {with} both gas and pebbles {contributing to the opacity}. At the end of the formation phase the young 10\me planet is embedded in the protoplanetary disk at 0.2AU, and its radius is the Hill radius (about 100\re). It contains 6.7\me of metal (silicate rock) and 3.3\me of gas (H,He). {More details on the planet formation model appear in appendix~\ref{app:form}}.\\

{\it Input parameters} are similar to what we used in the planet formation model: the equations of state of H, He \citep{scvh}, and $SiO_2$ rock \citep{faik18} {are the same} as in the formation work. Irradiation by the parent star is for the same location as in the disk phase (i.e., no migration). 
Since we don't expect grains or pebbles to remain in the envelope in the long term evolution \citep{brouwers21,movsh10,ormel14,mordasini14}, the radiative opacity is {that} of a grain-free solar metallicity atmosphere \citep{freedman14}.\\
    
{\it The thermal evolution model} {is calculated for the entire interior - from center to surface - on one mass grid, with no distinction between core and envelope. The model solves the interior structure and evolution equations (eq.~1-5 in \cite{vazan15}) which allows for heat transport by convection, radiation and conduction depends on local conditions in time (eq.~11-12 in \cite{vazan15}).
The redistribution of composition is by convective mixing where the convection criterion is fulfilled, and by rainout in oversaturated regions.
Convective-mixing is calculated as described in the appendix of \cite{vazan15}. In this work we add the simulation of composition redistribution by rainout, as explained in the next paragraph.}\\
    
{The} {\it Rainout model} simulates the condensation and settling of silicate vapor from an over-saturated layer into an undersaturated layer below, 
{similar to the approach of \cite{stevenson22} for modeling the evolution of Jupiter.}
For consistency with the formation models \citep{ormel21} we use the same liquid-vapor curve of \cite{kraus12} (lower branch) {to determine the saturation in the interior:
\begin{equation}
    \rho_{\rm sat}=\exp \left( a-\frac{b}{T}+\left(\frac{T}{T_{\rm cp}}\right)^c \right)
\end{equation}
where $a = 9.0945$, $b = 5.630\times 10^4$\,K, $c = 13.26$, and $T_{\rm cp} = 5130$\,K is the critical point of $SiO_2$. 
As the planet cools along the evolution, whenever the silicate mass fraction in a certain layer $Z_i$ exceeds the saturation level $Z_{\rm i,sat}$, we remove the excess of silicate (above the saturation level) to the layer below it. Numerically, if $Z_i>Z_{\rm i,sat}$ at time $t$ in layer $i$ the silicate mass fraction is:
\begin{equation}
      Z_i(t+1)=Z_i(t)-Z_{\rm i,sat}
\end{equation}
and the excess of rock $(Z_i-Z_{sat})m_i$ is removed from layer $i$ to the layer below it, enhancing its metallicity.}
The rock droplets are assumed to instantaneous fall between layers, as {the} droplets {settling} time is much shorter than the Kelvin-Helmholtz time. 
The energy of the rainout - {latent heat of} condensation and the settling energy - {are included in the evolution model via the equation of state and the interior structure equations respectively.}
{For simplicity, we ignore in this work the effect of convective inhibition in condensation front on slowing the heat transport {\citep{guillot95a,leconte17,markham22}.}\\

{\it Mass loss} is included in the thermal evolution model, by removal of the mass from the outermost layer of the planet {at} a constant rate. 
Since rainout of rock droplets from oversaturated layer is fast, mass loss (if {it} occurs) is of hydrogen and helium and not of silicates. 
The rate of mass loss by stellar XUV depends on assumptions on atmospheric absorption and stellar properties, and the range of possible outcomes is quite {diverse} \citep[e.g.,][]{lopez12,mordasini20,king21,kubyshkina22}. We therefore use constant mass loss rates to achieve {mass loss fractions} {amounting to} 0-0.9 {of the initial post-disk} envelope mass within the first 0.1\,Gyr, {during} which the stellar XUV is strong. 

\begin{figure}
\centerline{\includegraphics[width=9cm]{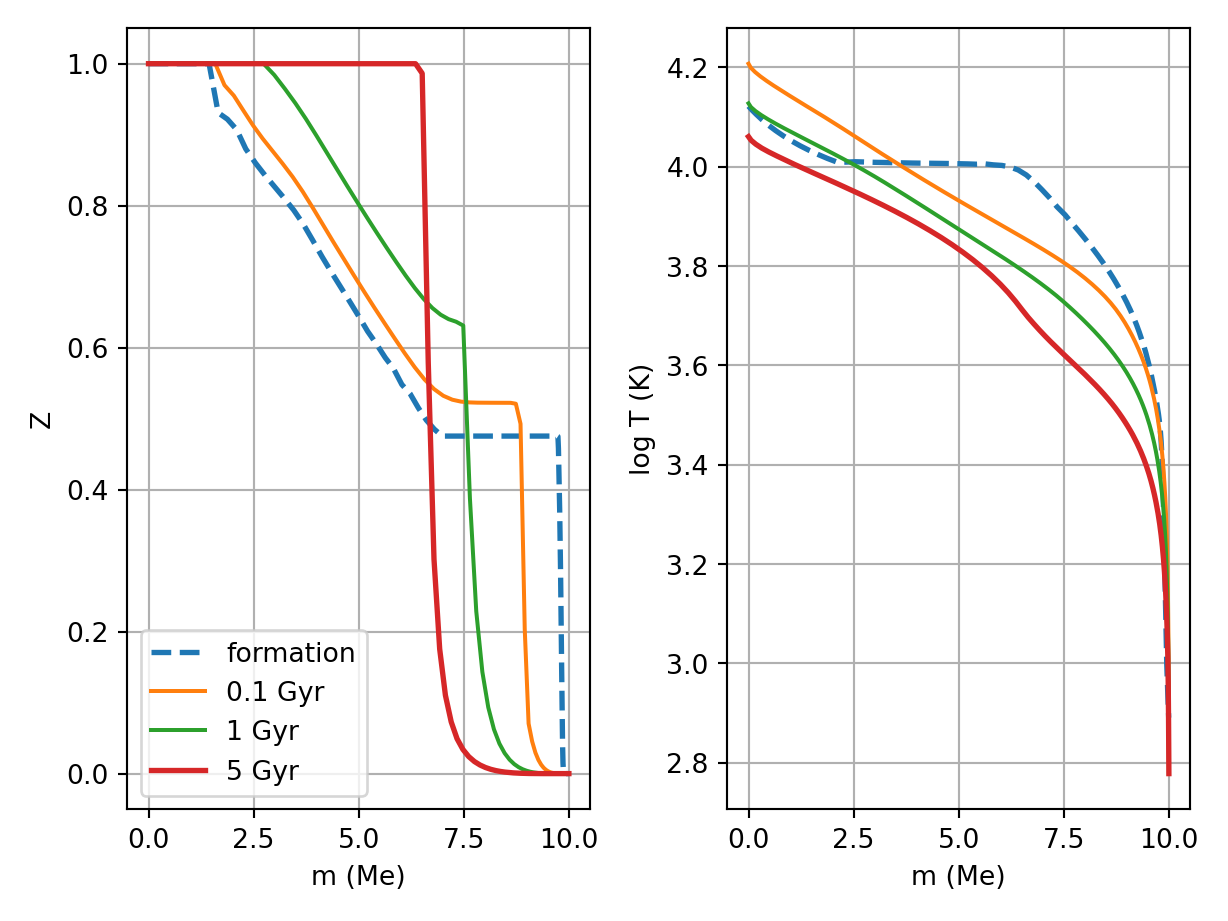}}
\caption{Silicate mass fraction (left) and temperature profile (right) of a 10\me sub-Neptune with 33\% H,He located at 0.2\,AU. Profiles are shown at different evolution phases. {See appendix~\ref{app:form} for more details on the initial (planet formation) profiles.} 
}\label{fig:ZTm}
\end{figure}
    
\section{Results}

\subsection{Sub-Neptunes with massive envelope from formation}

The planet formation phase {ends} with the dissipation of the disk and rapid contraction of the outer diluted gas layers from 100\re to about a tenth of {that size} within a few million years. Despite the significant radius {contraction}, the density and composition distribution in the deep interior is almost not affected.
After the rapid contraction, the long term cooling of the outer metal-rich envelope of this planet is governed by large scale convection, while the cooling of the deeper composition gradient is by conduction and/or layered convection, which is less efficient. Consequently, the metal-rich envelope cools faster and {the silicate vapor} becomes oversaturated, while the deep interior remains very hot and thus undersaturated. {As a result}, {metal} starts to {condense} and settle (rainout) into deeper layers.
As the planet cools the condensation {front} {moves} deeper in the interior. Since the liquid-vapor curve is a strong function of temperature, the layers above the condensation front are almost {metal}-free. 

\begin{figure}
\centerline{\includegraphics[width=9cm]{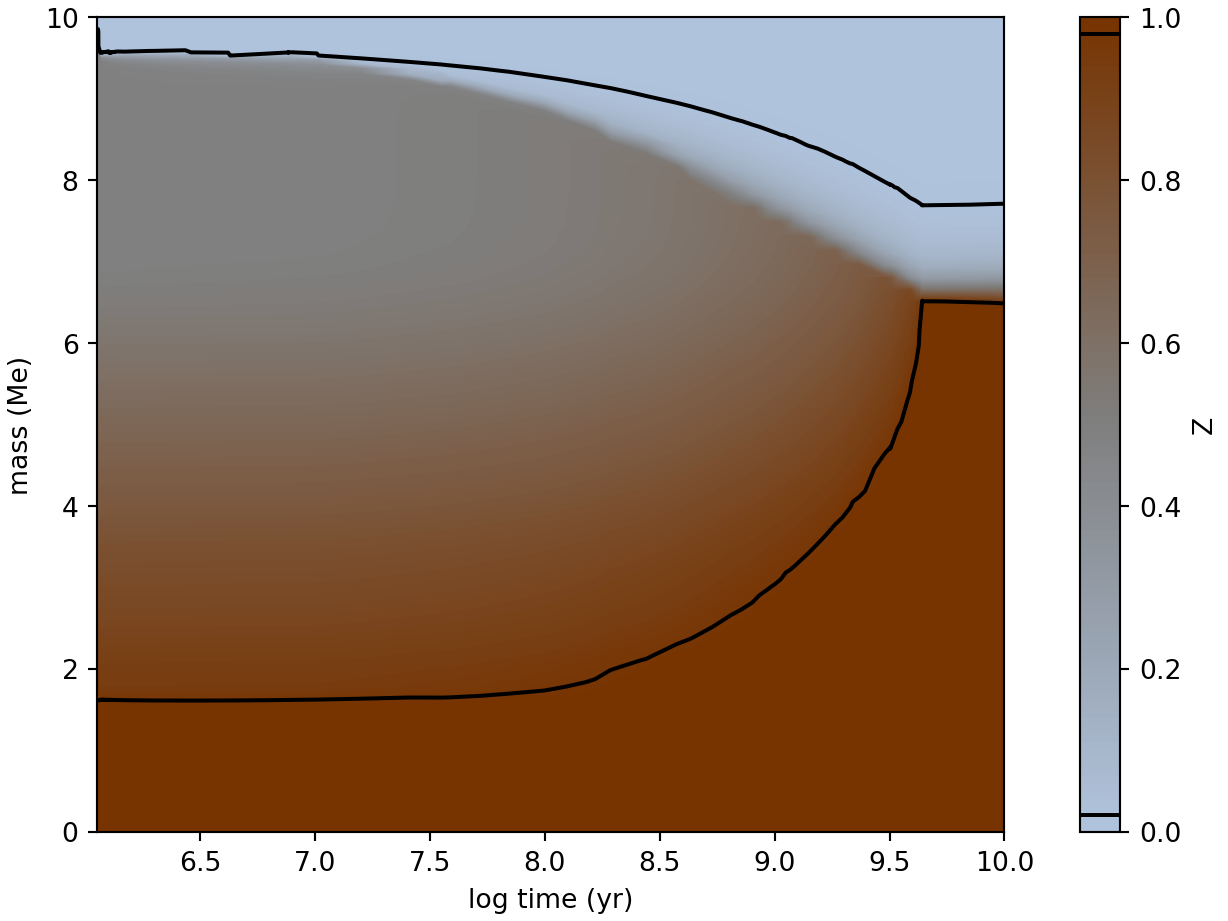}}
\caption{Silicate mass fraction (color) as a function of interior layers (y axis) and time (x axis). Silicate mass fraction ranges between zero (gas only) in blue and pure silicate in brown. The gradual distribution of silicate from formation converges into a core-envelope structure {after} about 4.25\,Gyr. The solid lines signify $Z=0.98$ and $Z=0.02$ enrichment {levels}. {Versions of this figure for radius and pressure layers instead of mass are presented in appendix~\ref{app:fig2}.}}\label{fig:Zmt}
\end{figure}
    
We find that the effect of rainout on the structure of rocky sub-Neptunes formed by pebble accretion is {significant}. In figure~\ref{fig:ZTm} we show the silicate (Z) mass fraction distribution (left) and temperature profile (right) of the 10\me planet located at 0.2AU. The initial state, based on the results of the formation model (dashed) and after 0.1, 1, and 5\,Gyr (solid) are shown. Although the planet's interior at the end of the formation phase contains most of its silicate as vapor in the polluted envelope, after 5\,Gyr all the silicate vapor {has} condensed and settled and the planet has {transitioned to} a core-envelope structure. 

Interestingly, the redistribution of silicate in the planet interior is found to take place on {billion} years timescale, as is shown in figure~\ref{fig:Zmt}. The early fast contraction increases the temperate at the outer layers in the first 10 Myrs, and therefore significant rainout of silicate starts only after about 20 Myr. The rainout continues as the planet cools and ends when all the silicates are located in a rocky core, at age of 4.25\,Gyr for this 10\me sub-Neptune. The simulation shown in figure~\ref{fig:Zmt} is for planet with its initial gas content of 33\% in mass (no mass loss). Since the H,He amount may decrease after disk dissipation, we will examine in section~\ref{sec:Menv} how the rainout timescale changes with envelope mass. 

{The rainout} releases energy {from the} condensation {process} (latent heat) and gravitational settling. 
Moreover, {the settling erases parts of the original composition gradient, resulting in the} release of formation energy that was stored {there, a consequence of low heat transport efficiency in} {regions characterized by} composition gradients. 
These energy sources cause {the planet radius to expand for} {over the duration of} the rainout period. When the rainout {terminates} the three energy sources {vanish} and the efficient (convective) cooling results in faster radius contraction. In the left panel of figure~\ref{fig:RLt} we present a comparison between the radius of a raining-out planet (red) formed by pebble accretion to a similar planet with static structure, i.e., no rainout\footnote{The model of rocky plant with polluted envelope and no rainout is unrealistic, and is shown here in order to emphasis the rainout effect on radius evolution.} (dashed black) and to a similar planet with core-envelope structure from formation (solid black). As is shown, the effect of rainout on radius evolution is significant. The luminosity of the planets is shown in the right panels, indicating significant luminosity increase by the rainout. 

\begin{figure}
\centerline{\includegraphics[width=9cm]{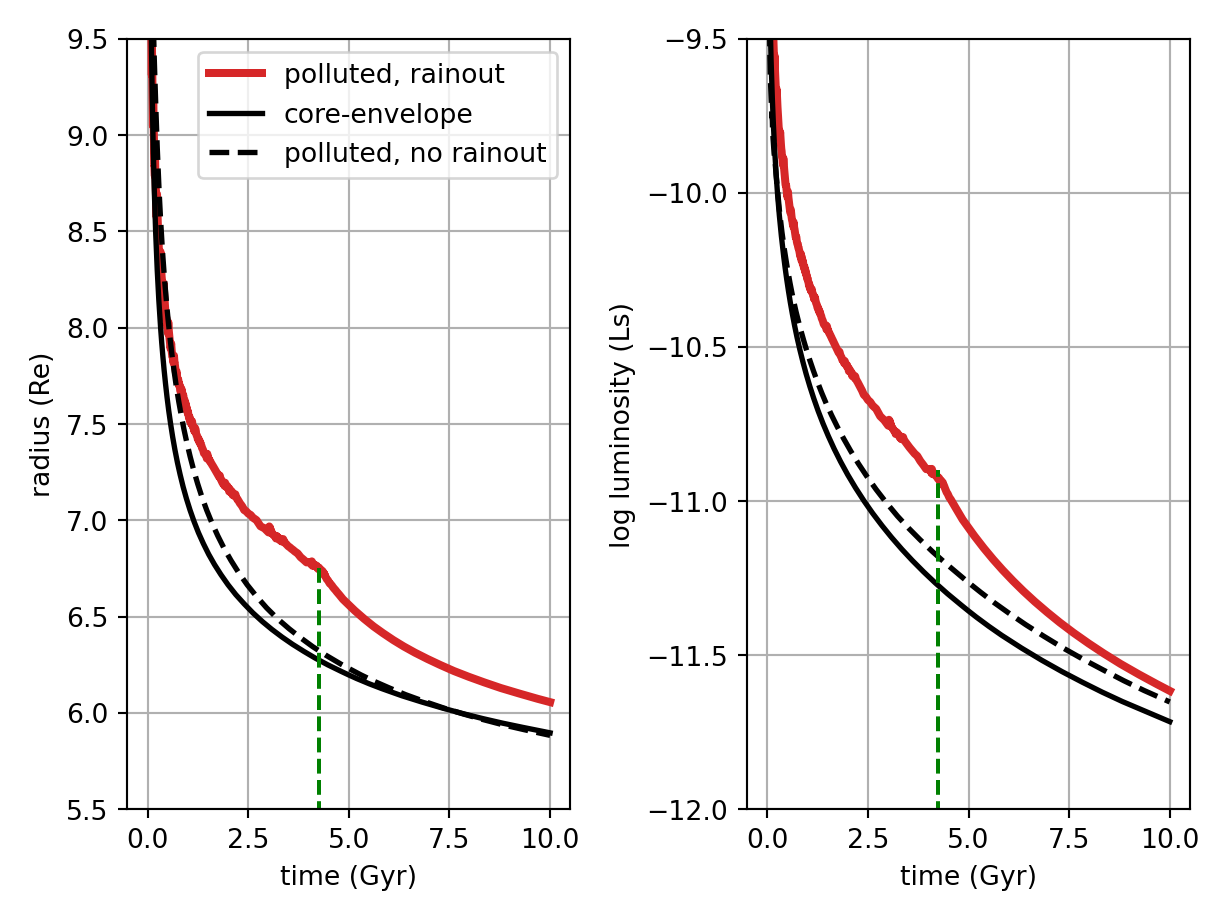}}
\caption{Radius (left) and luminosity (right) evolution of the 10\me planet shown in figures~\ref{fig:ZTm}-\ref{fig:Zmt} (red). The evolution of same planets without rainout (dashed black) and with core-envelope structure from formation (solid black) are shown for comparison. The green vertical line signifies the end of the rainout phase. Radius inflation {at} that point is 7.8\% in comparison to the core-envelope case. 
}\label{fig:RLt}
\end{figure}

We define {the} {\it rainout timescale} as the time from formation until all the silicates {have} settled and a core-envelope structure {has been} achieved.
We define {\it maximum radius inflation} as the radius change at the end of the rainout phase, in comparison to a similar (mass and composition) planet formed with core-envelope structure at the same age. 
In Fig.~\ref{fig:RLt} the end of the rainout phase, i.e., when the planet reaches a core-envelope structure, is illustrated by the green vertical dashed line. 
The rainout timescale is $t=4.25$\,Gyr and the maximum radius inflation is $\Delta R/R =0.078$. 
As can be seen, the slope of the radius evolution changes since the planet become fully adiabatic and cools more efficiently from that point. 

The radius inflation of the planet in Fig.~\ref{fig:RLt} is retained in the post-rainout phase. The energy release from massive envelopes is slow and the radius evolution is therefore delayed in comparison to planets formed with core-envelope structure.
In other words, planets that formed with polluted massive envelopes would look "younger" than planets that formed by core-envelope, caused by the later rainout heat release. 
The radius of the {raining-out} planet in Fig.~\ref{fig:RLt} is 2.5\,Gyr younger along the post-rainout evolution, than a similar planet that formed with a core-envelope structure. 
{Alternatively,} if age is known this larger radius might be inferred as a higher H,He content for such observed planet, {as will be discussed in the next section}.

\subsection{Sub-Neptunes with lower mass envelopes}\label{sec:Menv}

The duration of the rainout phase depends on the planet's cooling efficiency, hence on the blanketing effect of the envelope. Thicker envelopes will {retain} interior heat for longer. In some cases the initial envelope (or part of it) is lost after the disk dissipates by XUV radiation from the star. The sub-Neptune shown in figure~\ref{fig:Zmt} contains 33\% of H,He from formation, but a large fraction of this envelope can be removed in close-in sub-Neptunes, up to complete loss \citep[e.g.,][]{owenwu13,rogersowen21}. 

We next explore the rainout process and its {duration} {in simulations that include} mass loss{, which we express in terms of} the {remaining} envelope mass. 
We repeat the thermal evolution calculation for the same initial 10\me planet with different mass loss rates during the first 0.1\,Gyr of their evolution. In figure~\ref{fig:Menv} we present {by the green line} the rainout timescale as a function of the remaining H,He mass. As expected, the smaller the gas mass {(the greater the mass loss)} the faster the rainout. Interestingly, sub-Neptunes with {with H,He amounts exceeding 0.75\me} are in the rainout phase for billions of years{. They are thus characterized by} {inflated} radii and gradual / polluted interiors. Gas-poor sub-Neptunes with less than 0.75\me of H,He "converge" to core-envelope structure within {less than} 1\,Gyr, resulting in fast (adiabatic) cooling and contraction {in the post-rainout phase.
Yet, planets with envelope masses larger than 0.3\me would look "younger" by their late released heat content for Gyrs, while planets with less than 0.3\me of H,He cool fast enough that their radius evolution coincides with the radius evolution of a similar planet that was born with a core-envelope structure in 1-2\,Gyr.}
In that sense, rocky planets {with less than 0.3\me H,He} that formed with polluted envelopes cannot be distinguished from planets that formed with a core-envelope structure after $\sim$1\,Gyr.

\begin{figure}
\centerline{\includegraphics[width=\columnwidth]{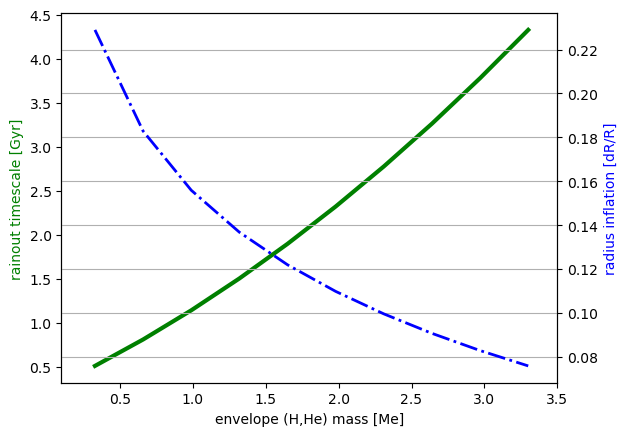}}
\caption{{Green}: time from formation until convergence to a core-envelope structure (rainout timescale) as a function of envelope mass. Trend is shown for sub-Neptune planets that contain 6.7\me of silicates and their gas (H,He) mass is determined by the mass loss rate, from 3.3\me (original model, no mass loss) down to 0.33\me {($3\,M_\oplus$ mass loss)}. 
{Blue}: maximum radius inflation by rainout in comparison to the core-envelope structure model {at the rainout timescale}. Curves are polynomial fits for the evolution data points.}\label{fig:Menv}
\end{figure}

Faster rainout in planets with low mass envelopes causes {stronger} radius inflation, as similar amount of energy is being released in shorter time. Thus, the $\sim$8\% radius inflation of a planet with massive (3.3\me) envelope at age of 4.25\,Gyr becomes {23\%} radius inflation in a planet with 0.33\me envelope at 0.55\,Gyr. 
In figure~\ref{fig:Menv} we present (dashed blue) the trend of maximum radius inflation with {remaining} envelope mass.
As can be seen, the earlier the rainout occurs the larger the radius inflation is. 
The early but stronger radius inflation in planets with low H,He mass might enhance mass loss by photoevaporation, which depends on the planetary radius, if it overlaps with the strong XUV phase. 
The longer and moderate radius inflation in planets with significant H,He envelopes can be inferred, when using core-envelope model, as a larger H,He content.

\section{Implications for exoplanets}
The envelope (H,He) {mass} fraction of observed planet{s are} usually derived from mass-radius relations, {assuming a} core-envelope structure  \citep[e.g.,][] {lopezfor14,chenrog16,baraffe08}. 
{However, when inflation by rainout operates the H,He fractions inferred for these planets in this way would be overestimated.} 
To understand the significance of rainout, we examine a selection of planets in the exoplanet census with  well defined masses, radii, and ages.  
Since the radii of these planets are more sensitive to their H,He fraction than to their planetary mass \citep{lopezfor14}, we extrapolate our results to slightly larger and smaller planets {between 7-12\me}.
{For each exoplanet we derived the core-envelope H,He fraction based on \cite{lopezfor14} for standard (solar) opacity. We then calculated the expected radius inflation by rainout at its age, and used it to correct the radius and obtain a new H,He fraction, taking into account the rainout {effect}. Finally, we run models with the new H,He fraction to verify {their} consistency with the observed parameters. As can be seen in Table~\ref{tab:exop}, the actual H,He content can be half of the estimated by the standard model, {when} these planets formed with polluted envelopes.}

\setlength{\tabcolsep}{10pt} 
\renewcommand{\arraystretch}{1.3} 
\begin{table*}[]
    \centering
    \begin{tabular}{lcccccc}
    \hline
    \hline
     & mass & radius & period & age & H,He  & H,He \\
     & (\me) & (\re) & (day) & (Gyr) & standard & w/ rainout \\
     \hline
     Kepler-36 c & 7.13 &  3.6 &  16.2 & 6.9{\footnotesize $\pm 0.372$} & 7.8\% & 6\% \\
     Kepler-11 e & 7.95 &  4.1 &  32 & $8.5^{+ 1.1}_{- 1.4}$ & 15\% & 11\% \\
     TOI-1136 d & 7.95 & 4.53  & 12.5  & 0.7{\footnotesize $\pm 0.1$} & 13\% & 6.5\% \\
     TOI-1136 f & 8.3 &  3.8 &  26.3 & 0.7{\footnotesize $\pm 0.1$} & 9\% & 4.5\% \\
     TOI-1422 b & 8.9 &  3.88 &  13 & $5.1^{+ 3.9}_{- 3.1}$ & 10\% & 6\% \\
     KOI-142 b & 9.54 &  3.37 &  10.9 & $2.4^{+ 1.2}_{- 0.77}$ & 7\% & 4\% \\
     K2-314 d & 10.2 & 3.86  & 35.7  & 9{\footnotesize $\pm 0.6$} & 13\% & 10\% \\
     K2-19 c & 10.8 &  4.76 &  11.9 & 8+ & 20\% & 14\% \\
     Kepler-79 b & 10.9 &  3.4 &  13.5 & $3.4^{+ 0.6}_{- 0.91}$ & 6.6\% & 4.5\% \\
     Kepler-30 b & 11.4 &  3.82 & 29.3  & 2{\footnotesize $\pm 0.8$} & 11\% & 6\% \\
     \hline
    \end{tabular}
    \caption{Parameters of observed exoplanets {\it (exoplanet.eu)} that we used {as input parameters} in the model, and the resulting H,He fraction estimates for standard core-envelope models and for planets formed with polluted envelopes and rainout their silicates {afterwards}. The selected planets are all the exoplanets with masses 7-12\me, radii above 3.3\re, and estimates for stellar age {(marked in green in Figure~\ref{fig:MR}). Error bars were omitted from all observation values and we calculate each model for a single set of values.}}
    \label{tab:exop}
\end{table*}


In Figure~\ref{fig:MR} we show the mass-radius relation of observed exoplanets, and mark (in green) the planets {that} we model and are potentially affected by the rainout, either by post-rainout radius inflation or {because they are still} raining-out their polluted envelopes. 
{Planets with radii smaller than 3.3\re have less than 5\% H,He, thus the rainout effect after 1\,Gyr is negligible.}
Sub-Neptunes with radii above 3.3\re may have underestimated metal content if rainout effect on the radius is ignored, as is shown in Table~\ref{tab:exop}. Planets above 3.3\re that aren't marked in green don't have well defined age estimate and therefore cannot be interpreted by our model. \\

\begin{figure}
\centerline{\includegraphics[width=9cm]{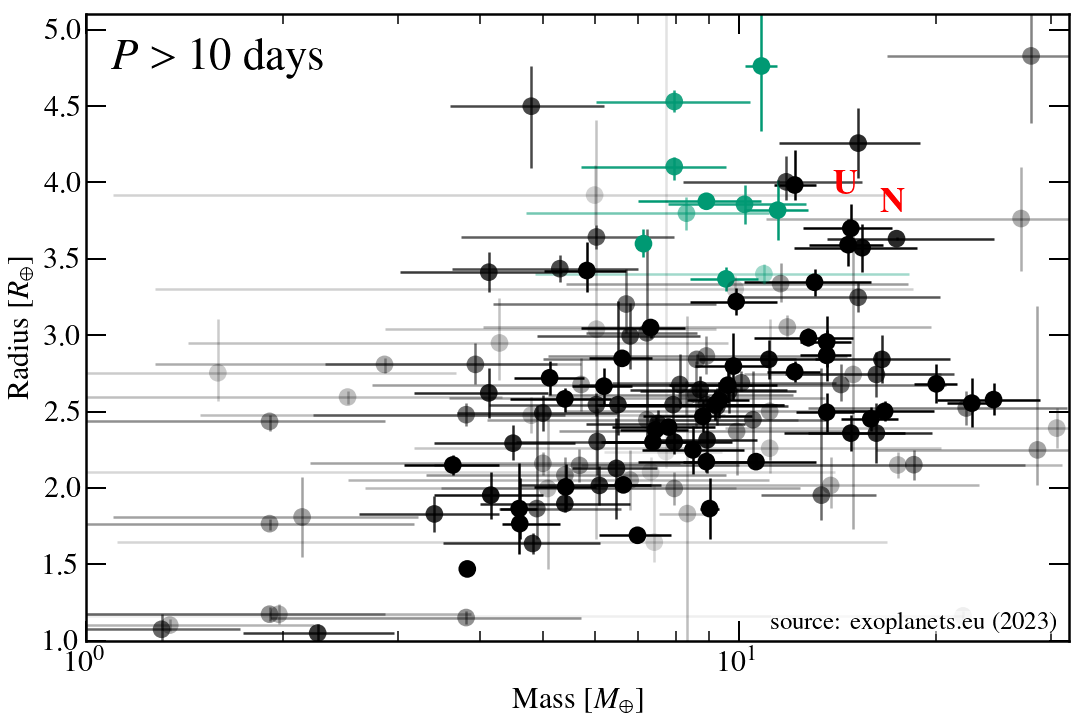}}
\caption{Mass-radius relation of observed exoplanets. Planets in green are potentially inflated by past or ongoing rainout, in comparison to planets that formed with core-envelope structures. The rainout effect on the inferred H,He mass of the marked exoplanets appear in Table~\ref{tab:exop}. {The sub-Neptune planets with radii above 3.3\re that aren't marked in green lack an age measurement and thus are not included in our analysis. Planets with radii below 3.3\me are consistent with low ($<5$\%) H,He and thus rainout has negligible effect on radius at Gyrs age.}}\label{fig:MR}
\end{figure}


{Rainout takes place on ${\sim}$Gyr timescale, making the age of the} planet an important parameter in determining its planetary structure. 
As can be seen from Table~\ref{tab:exop}, younger sub-Neptunes are prone to larger rainout effects. 
Since sub-Neptunes rain out on {billion }year timescale, which is the age of most of the observed exoplanets, knowing their age will help us to better constrain the interior structure of individual observed planets. As of today, many sub-Neptune planets don't have well defined age measurement. Future measurements of stellar age will allow us to suggest corrected H,He fractions for these planets, for the case they formed with polluted envelopes.  
Moreover, large enough statistical database of radius and age will allow us to examine trends in radius evolution which may hint towards formation mechanisms of sub-Neptunes. For example, planets formed with polluted envelopes will feature {a spread in radius during their} early evolution which is larger than expected by standard adiabatic cooling. 
Planets may {display} a typical radius diversity for a specific range of H,He mass content. 
In this context, the expected data from the PLATO mission \citep{rauer14short} on stellar age is expected to drive a significant improvement in our understanding of planets evolution in general, and specifically of planets formed with polluted envelopes and on the rainout process. \\

Another {observational implication} is the possible overlap of the rainout phase with the period of high energy irradiation by the young star. 
Since mass loss by photoevaporation strongly depends on planetary radius (third power, \cite{owen16}), planets with low-mass envelopes and thus early rainout and large radius inflation, are prone to enhanced "sedimentation-powered" mass loss. As we find, a 7\me sub-Neptune planet with 5\% H,He content will suffer radius inflation of up to 23\%, within the first 0.5\,Gyr, which might double the expected mass loss by photoevaporation at early time. 
Thus, planets with less than 0.3\me may experience "sedimentation-powered" mass loss of their primordial hydrogen atmospheres. 
A detailed calculation of the effect of radius inflation on mass loss will be discussed in a companion study (Vazan et al. - in prep.)

\section{Conclusions}

In this work we find that rainout {of high-Z vapor} {significantly affects the} evolution of sub-Neptunes born with polluted envelopes {as it can operate on evolutionary ($\sim$Gyr) timescales}. 
Timescales we find in this work are {nevertheless} shorter than in \cite{brouwers20}, which estimated 10\,Gyr rainout time for super-Earth planets with a few percent H,He. 
The longer timescale found in their work is probably the result of relying on the ideal gas equation of state, which usually suffers from too high temperatures for the same conditions \citep[see][for equation of state comparison]{ormel21}.

The {key findings from this study} {are as follows }:
\begin{enumerate}
    \item Sub-Neptune planets formed with polluted envelopes {experience} late growth of a rocky core {due to} rainout {of high-$Z$ vapor}. {The} timescale for rainout depends on envelope mass. 
    
    \item The interior structure of a rocky sub-Neptune formed by pebble accretion depends on its envelope (H,He) mass and age. Planets with less than 0.75\me H,He {evolve towards a} core-envelope structure {by} 1\,Gyr, while planets with more massive envelopes or younger planets may still be raining-out {when} we observe them, and thus have interiors with composition gradients and/or polluted envelopes.

    \item Rocky planets with less than 0.3\me H,He that formed with polluted envelopes cannot be distinguished from planets that formed with a core-envelope structure after about 1\,Gyr, due to short rainout phase and efficient cooling. 
    
    \item The Rainout process causes radius inflation by release of gravitational energy (settling), latent heat (condensation), and formation energy (composition gradient erosion). The radius inflation is shorter and larger in planets with low mass envelopes: a few percent for sub-Neptunes with massive gas envelope, and larger (tens of percent) but on much shorter timescale in planets with lower gas mass fraction. 
    
    \item Planets that formed with polluted envelopes would look "younger" than planets that formed with core-envelope structure, as a result of the later heat release. Alternatively, if age is known, the inflated radii might be interpreted as higher H,He mass in these planets, which can be up to twice of the actual value. 
    
\end{enumerate}

The model in this Letter is for rocky sub-Neptunes, formed as 10\me planet, with and without envelope mass loss. Significantly more (less) massive planets have stronger (weaker) gravity fields, which will affect the results quantitatively. In a companion paper (Vazan et al. - in prep.) we execute parameter space study for a wider range of masses and formation parameters.
  
\section*{Acknowledgments}
We thank the referee for constructive comments that improved the manuscript. A.V. acknowledges support by ISF grants 770/21 and 773/21.  C.W.O. acknowledge support by the National Natural Science Foundation of China (grant no. 12250610189). Figures were plotted using Matplotlib \citep{matplotlib07}. 

\bibliographystyle{aa}
\bibliography{allona} 

\begin{appendix}

\section{Planet formation model}\label{app:form}

The input model we use for the evolution is the outcome of the planet formation model of \cite{ormel21}, which calculates the growth of the core and envelope of an embedded, pebble-accreting planet. 
The envelope is assumed to be in hydrostatic equilibrium on dynamical timescales and in pressure equilibrium with the surrounding disk gas. At each time step a steady solution to the stellar structure equations is found. A key element in the model is the relaxation of the standard assumption that accreting solids end up in the core. 

The formation calculation assumes constant solid accretion rate of $10^{-5}${\me\,yr$^{-1}$}. The effects of pebble accretion on gas accretion are included via acceleration of gas accretion through accumulation of a vapor-rich atmosphere, and by contribution to the opacity, which reduces the intake of nebular gas. 
The formation model uses tabular equation of states of H,He and $SiO_2$ and their mixtures, based on \cite{vazan13}. The vaporization is determined by the silicate liquid-vapor (saturation) curve of \cite{kraus12}.

In figure~\ref{fig:form} we show the initial model we use in this work. The model is identical to what appears in the middle-bottom panel of figure~13 in \cite{ormel21}.
It is the outcome of the growth of a planet embedded in the protoplanetary disk at 0.2\,AU, starting from an embryo of 0.06\me and proceeding to the point where 10\me mass is reached after about 5\,Myr. 
{The solid core in the planet formation model is not part of the interior grid and is modeled by constant density and arbitrary temperature. These input density and temperature profiles converge to isentropic profiles after the initial procedure in the evolution model, in which realistic equation of state and structure equations are applied also to the core.}

{The end of the planet formation phase is chosen at the state where the planet mass reached 10\me. 
As expected, if the simulation is allowed to continue in gas-rich vicinity the planet would reach the crossover point and start its runaway gas accretion. Nevertheless, pre-runaway evolutionary timescales typically slow down \citep[e.g.,][figure~6]{rafikov06,leechiang15,ormel21}, making the formation of such 10\me planet with a significant H,He envelope conceivable.}

\begin{figure}
\centerline{\includegraphics[width=8cm]{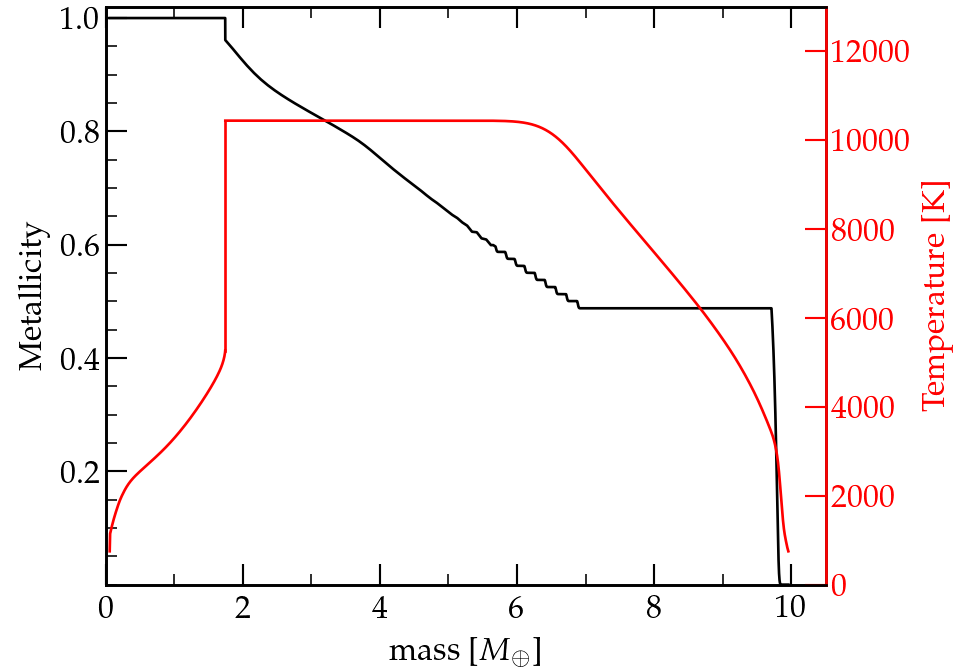}}
\caption{The initial model we use for the evolution calculations, resulting from planet formation model of \cite{ormel21}. 
Metallicity (black) and temperature (red) profiles for a planet formed by rocky pebble accretion at 0.2\,AU. The temperature in the classical core region that is not modeled in the formation model, is fitted by isentropic profile in the evolution model, see text for details. 
}\label{fig:form}
\end{figure}

\section{Rainout vs. radius and pressure}\label{app:fig2}

{The data shown in figure~\ref{fig:Zmt} as a function of the code's mass grid (y-axis) may give the impression that rainout takes place in the outer atmosphere. We therefore plotted the same data also as a function of radius (figure~\ref{fig:Zrt}) and pressure (figure~\ref{fig:Zpt}) in the y-axis. Solid lines indicate values of $Z=0.02$ and $Z=0.98$.
As can be seen in the two figures, the envelope pollution (evaporation front) is located in the deep interior at pressure $>0.1$\,GPa, at about half of the total radius. While rainout takes place deep in the envelope, the outer envelope (and atmosphere) are left almost silicate-free, and thus prone to easier inflation by the rainout heat release. }

\begin{figure}
\centerline{\includegraphics[width=9cm]{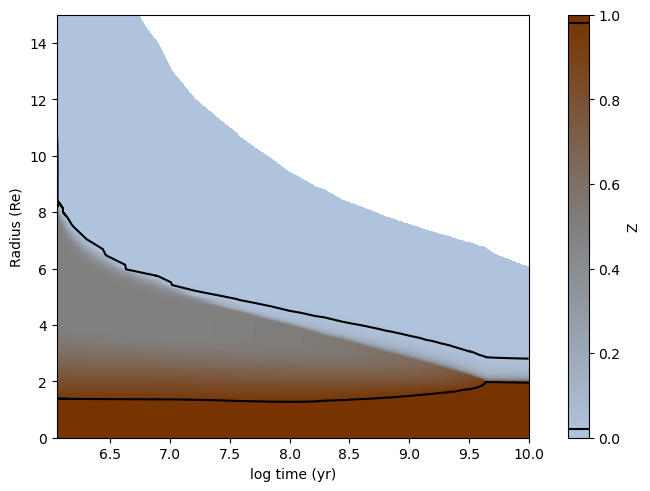}}
\caption{Same data of figure~\ref{fig:Zmt}, here as a function of radius instead of mass (y axis). Silicate mass fraction (color) ranges between zero (gas only) in blue and pure silicate in brown. White area is empty, due to radius contraction.}\label{fig:Zrt}
\end{figure}

\begin{figure}
\centerline{\includegraphics[width=9cm]{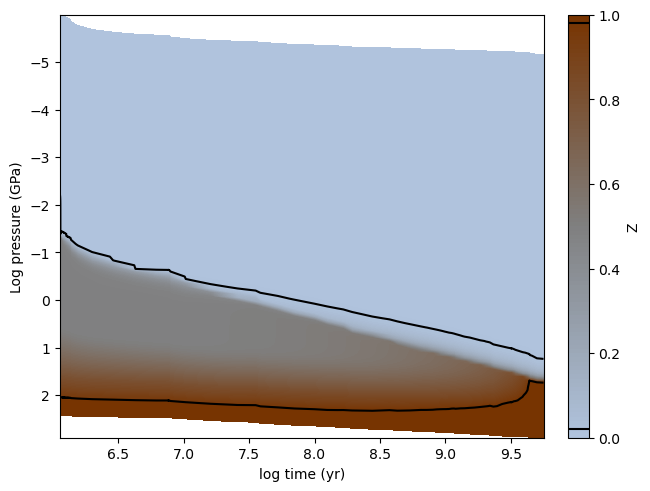}}
\caption{Same data of figure~\ref{fig:Zmt} and~\ref{fig:Zrt}, here as a function of pressure (y axis). Silicate mass fraction (color) ranges between zero (gas only) in blue and pure silicate in brown. White areas are empty, due to pressure increase in contraction.}\label{fig:Zpt}
\end{figure}

\end{appendix}

\end{document}